\documentclass[prl,showpacs,reprint,floatfix, groupaddress,superscriptaddress,footinbib,longbibliography]{revtex4-1}
\usepackage{times,amsmath,amsfonts,amssymb,latexsym}
\usepackage{graphicx,epsf}
\usepackage{subfigure}
\usepackage{cancel}

\usepackage{color}
\newcommand{\be}{\begin{equation}}
\newcommand{\ee}{\end{equation}}
\newcommand{\ba}{\begin{eqnarray}}
\newcommand{\ea}{\end{eqnarray}}

\newcommand{\ignore}[1]{}

\def\CC{{\rm\kern.24em \vrule width.04em height1.46ex depth-.07ex
    \kern-.30em C}}
\def\P{{\rm I\kern-.25em P}}
\def\RR{{\rm
         \vrule width.04em height1.58ex depth-.0ex
         \kern-.04em R}}

\def\bbbc{{\mathchoice {\setbox0=\hbox{$\displaystyle\rm C$}\hbox{\hbox
to0pt{\kern0.4\wd0\vrule height0.9\ht0\hss}\box0}}
{\setbox0=\hbox{$\textstyle\rm C$}\hbox{\hbox
to0pt{\kern0.4\wd0\vrule height0.9\ht0\hss}\box0}}
{\setbox0=\hbox{$\scriptstyle\rm C$}\hbox{\hbox
to0pt{\kern0.4\wd0\vrule height0.9\ht0\hss}\box0}}
{\setbox0=\hbox{$\scriptscriptstyle\rm C$}\hbox{\hbox
to0pt{\kern0.4\wd0\vrule height0.9\ht0\hss}\box0}}}}
\def\bbbz{{\mathchoice {\hbox{$\sf\textstyle Z\kern-0.4em Z$}}
{\hbox{$\sf\textstyle Z\kern-0.4em Z$}}
{\hbox{$\sf\scriptstyle Z\kern-0.3em Z$}}
{\hbox{$\sf\scriptscriptstyle Z\kern-0.2em Z$}}}}

\usepackage{hyperref}

\begin{document}

\preprint{APS/123-QED}

\title{Direct observation of magnetic monopole freedom in two-dimensional artificial spin ice}

\author{D. G. Duarte}
\affiliation{Laborat\'{o}rio de Spintr\^{o}nica e Nanomagnetismo, Departamento de F\'{i}sica, Universidade Federal de Vi\c{c}osa, Vi\c{c}osa,
36570-900, Minas Gerais, Brazil}
\author{L. B. de Oliveira}
\affiliation{Laborat\'{o}rio de Spintr\^{o}nica e Nanomagnetismo, Departamento de F\'{i}sica, Universidade Federal de Vi\c{c}osa, Vi\c{c}osa,
	36570-900, Minas Gerais, Brazil}
\author{F. S. Nascimento}
\affiliation{Grupo de F\'{i}sica, Universidade Federal do Rec\^{o}ncavo da Bahia, 45300-000 - Amargosa - Bahia - Brazil.}
\author{W.A. Moura-Melo}
\affiliation{Laborat\'{o}rio de Spintr\^{o}nica e Nanomagnetismo, Departamento de F\'{i}sica, Universidade Federal de Vi\c{c}osa, Vi\c{c}osa,
	36570-900, Minas Gerais, Brazil}
\author{A. R. Pereira}
\affiliation{Laborat\'{o}rio de Spintr\^{o}nica e Nanomagnetismo, Departamento de F\'{i}sica, Universidade Federal de Vi\c{c}osa, Vi\c{c}osa,
	36570-900, Minas Gerais, Brazil}
\author{C. I. L. de Araujo}
\affiliation{Laborat\'{o}rio de Spintr\^{o}nica e Nanomagnetismo, Departamento de F\'{i}sica, Universidade Federal de Vi\c{c}osa, Vi\c{c}osa,
	36570-900, Minas Gerais, Brazil}

\date{\today}

\begin{abstract}
Magnetic monopole unpairing as a function of external magnetic fields is presented as a fingerprint of this emergent quasiparticles freedom in a two-dimensional artificial spin ice system. Such freedom, required for example for further application in magnetricity, is only possible due to ground-state degeneracy, which causes a decreasing of the string energy in rectangular geometries, designed to allow highest equidistance among nanomagnets. We show by simulations that spin correlation in different rectangular artificial spin ices evolves from antiferromagnetic ordered magnetic structure to a ferromagnetic one, passing through an ice regime were pinch points related to Coulomb phase are observed. By measurements of magnetic force microscopy, we observe magnetic monopole creation, transport and annihilation in such systems with free monopoles created and transported throughout the sample without strings attached, as is commonly observed in conventional artificial spin ice systems.

\end{abstract}

\maketitle
Artificial spin ices (ASI), originally composed by Permalloy nanomagnets arranged in a square lattice \cite{wang2006artificial}, were proposed as a suitable system yielding magnetic monopoles due to geometrical frustration \cite{mol2009magnetic}. Direct observation of such emergent quasiparticles at room temperature was subsequently accomplished by magnetic force microscopy \cite{ladak2010direct}, despite the predicted ground state obeying ice rule at each vertex has often escaped from observation in real samples. In ASI's such monopoles are nowadays termed Nambu-like monopoles \cite{silva2013nambu,morley2019thermally}, once they are connected by energetic strings, as it takes place with their high energy counterparts\cite{nambu1974strings}.

The reasonable simplicity in the fabrication and manipulation of such artificial systems has led to proposals for utilizing magnetic monopoles in information technology, once their flow according to applied magnetic field generates magnetricity \cite{bramwell2009measurement,loreto2015emergence, goncalves2019tuning}, as much like electric field yields electricity in usual circuits. However, the energetic strings connecting monopole pairs in a square ASI deeply jeopardizes such an application, since when these pairs move apart a wake of energetic strings is left behind affecting the motion of other monopoles.

Inspired by freely moving monopoles in natural crystals of spin ice, it was predicted that ASI may also lead to monopole freedom when a critical height offset between nanomagnets layer is accomplished \cite{moller2006artificial,mol2010conditions}. Only very recently, experimental realization of such a tricky geometry was achieved with the fabrication of three-dimensional (3D)  ASI \cite{perrin2016extensive}, in which magnetic charges propagate as monopole plasma thermally excited \cite{farhan2019emergent}. In addition, free monopoles moving under external magnetic field has been recently observed in a macroscopic counterpart of those 3D ASI's\cite{teixeira2021macroscopic}. Despite the monopole freedom achieved in 3D ASI's, the fabrication process with additional steps of thin film deposition, milling definition of nanomagnets and spatial alignment needed, could prevent its utilization in micro and nanoscaled magnetricity device application. Another difficulty relies on the intrinsic distance non-uniformity between nanomagnet layers and the magnetic force microscope tip, a fact that may trouble its magnetic characterization.

In this letter, we report the first direct observation of monopole freedom under external magnetic field in a two-dimensional nanoscaled ASI. Actually, by stretching one side of the square lattice, a rectangular geometric arrangement is obtained. There, the interplay between degeneracy of ground state and geometrical frustration was predicted to ensure vanishing of string tension \cite{nascimento2012confinement,ribeiro2017realization,loreto2018experimental}. Such a fact is now observed and in this tensionless string framework magnetic monopoles move practically freely in a relatively ordered way.

\begin{figure}[!hbt]
    \centering
    \includegraphics[scale=0.285]{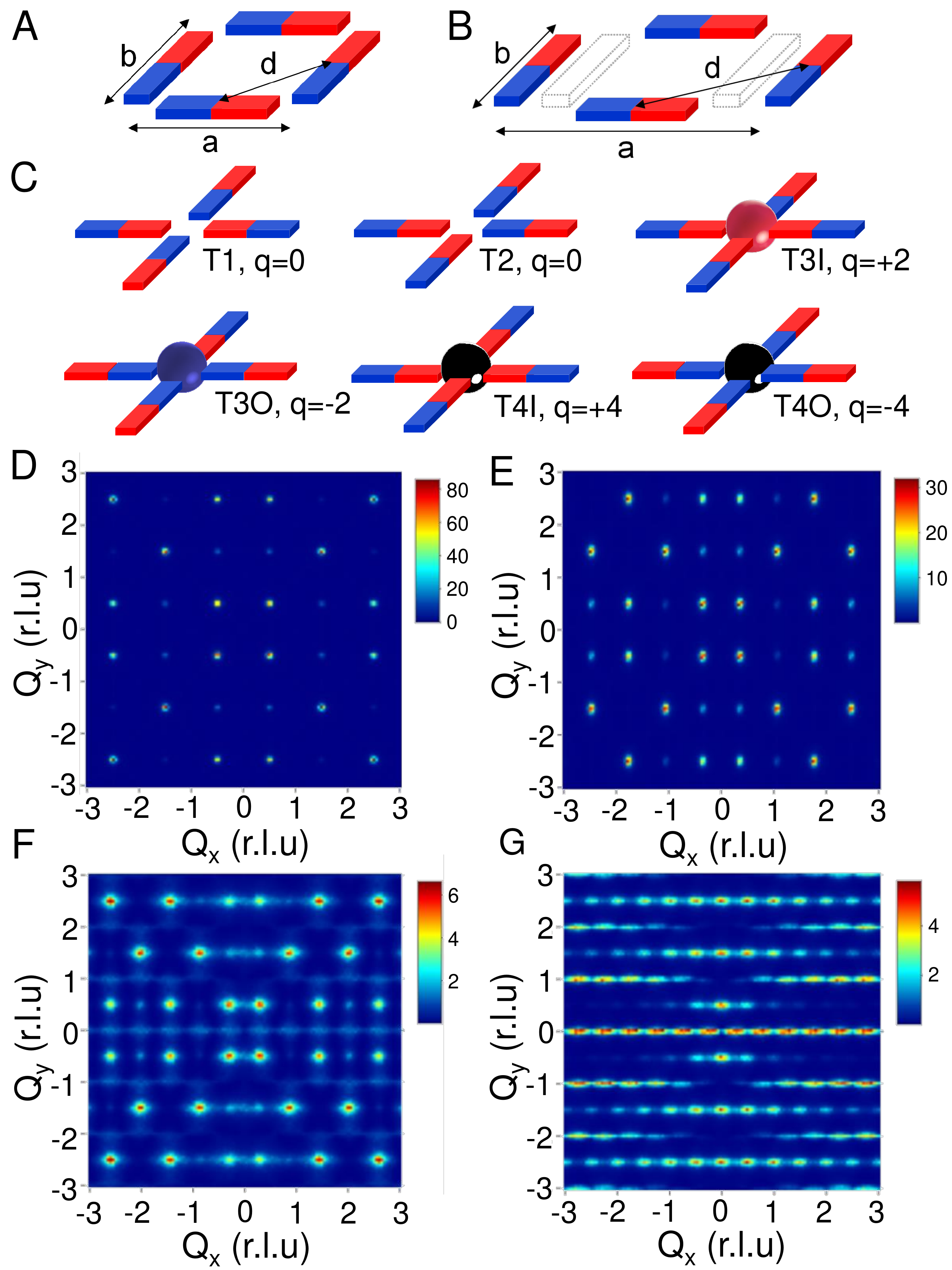}
    \caption{A - Conventional two-dimensional square geometry with lattice parameter $a=b$ and diagonal d, B - proposed rectangular lattice with $a > b$, C - vertex topology possible for all the geometries investigated with associated magnetic charges. Spin correlation obtained from dipole Monte Carlo simulation performed in D - conventional square lattice, E - two-dimensional rectangular lattice with $a=\sqrt{2}b$, F - two-dimensional rectangular lattice with $a=\sqrt{3}b$ and G - two-dimensional rectangular lattice with $a=\sqrt{4}b$.}
    \label{fig:fig1}
\end{figure}

Monte Carlo simulation of dipole-dipole interaction has been used to achieve thermal annealing in different rectangular lattices and also to calculate the magnetic structure factors (1000 lower energetic configurations have been tested; further details may be found in the Supplementary Material with low energy states for the distinct lattice configuration presented). Spin correlations have been obtained for square and rectangular arrangements, depicted in Figure 1A and 1B, respectively: square lattice has its diagonal larger than the equal sides $d>a=b$, while in the rectangular case, we have considered three distinct cases concerning its aspect ratio, $a=\sqrt2$, $a=\sqrt3$, and $a=\sqrt4=2$, referred as $R2$, $R3$, and $R4$ arrangements. Despite the aspect ratio, rectangular lattice vertices may be arranged into 6 topologies (Fig. 1C) which can be classified as vertices $T1$, $T2$, $T3I$, $T3O$, $T4I$ and $T4O$. $T1$ and $T2$ obey ice rule, 2-in 2-out, and they bear lower energy than $T3$-type vertices, which contains magnetic monopole of charge $+2$ ($T3I$) and $-2$ ($T3O$). $T4I$ and $T4O$ are vertices with very high energy and monopole charges $+4$ and $-4$, respectively. Note that in the R3 case $a>d=b$, which is the reason why degeneracy between $T1$ and $T2$ is observed (Figure 1B).

Square lattice presents ordered ground state where ice rule is obeyed (exception is observed at the borders; further details may be found in the Supplementary Material, see namely Figure 1 therein). Such an ordered configuration yields intense and well defined spots at the corners of the Brillouin zone, as illustrated by its magnetic structure factor in Figure 1D. Allowing $a>b$ we depart from square to the rectangular arrangement for the nanomagnets. Here, the aspect ratio, $a/b$, plays a crucial role: namely for $a=\sqrt3 b$ energy degeneracy between $T1$ and $T2$ vertices yields vanishing of string tension and consequently enables monopoles to move freely. $R2$ arrangement is still ordered with dominance of $T1$-vertex throughout the sample (see Figure 2 in the Supplementary Material for further details). Besides that, there is also a spreading and intensity lowering in the spots of magnetic structure factor when compared to the square lattice, as depicted in Figure 1E. Such a spreading may be attributed to 'residual charges' which takes place, even in the $T1$ vertices, due to the asymmetric proximity between nanomagnets in rectangular arrangements\cite{nascimento2012confinement}.

In the $R3$ case, $T1$ and $T2$ topologies are energetically degenerate and monopoles are now connected in pairs by tensionless strings \cite{nascimento2012confinement,ribeiro2017realization,loreto2018experimental}. In addition, $T1$ and $T2$ vertices appear in equal number while unpaired $T3O$ and $T3I$ are also observed (see Figure 3 in the Supplementary Material). Such a mismatch is claimed to yield scattered background of magnetic structure factor with pinch points related to the Coulomb phase \cite{henley2010coulomb, perrin2016extensive} (Figure 1F). Finally, in the $R4$ arrangement the nanomagnets belonging to vertical and horizontal axis are almost non-interacting and the system presents ferromagnetic alignment along a given line (stripe) and it is populated by paired monopoles (see namely Figure 4 in the Supplementary Material). However, parallel stripes couple each other in an antiferromagnetic way (as may be seen in the Figure 5D, Supplementary Material), rendering the spots presented in Figure 1G representing the two sub-lattices, like occurs in usual antiferromagnetic systems. 

Experimentally, we have constructed rectangular arrays of $R2$, $R3$, and $R4$ with sample dimensions of $100 \mu m \times 100 \mu m$. Each nanomagnet is fabricated as a trilayer 3nm tantalum / 20nm permalloy / 3nm tantalum defined on the top of commercial silicon substrate, with lateral dimensions $3\mu m \times 400nm$. Such a shape ensures each magnet to behave as a Ising-like spin, whose magnetization points along the larger axis. Tantalum layers are used for better adhesion and as a capping to prevent Permalloy oxidation. First, samples have been completely saturated by applied magnetic field in the positive direction of high axis (horizontal). Afterwards, images have been taken by magnetic force microscopy (MFM) at remanence, for each magnetic field step applied in order to saturate magnetization in the opposite negative direction. Other measurements have been performed for two different directions, low axis (vertical) and diagonal, in order to avoid possible direction influence on the results (see Supplementary Materials Figure 6). Our goal here is to direct observe emergent quasiparticle behavior during creation, transport and annihilation of magnetic monopoles as function of applied magnetic field. 

\begin{figure}[!hbt]
    \centering
    \includegraphics[scale=0.068]{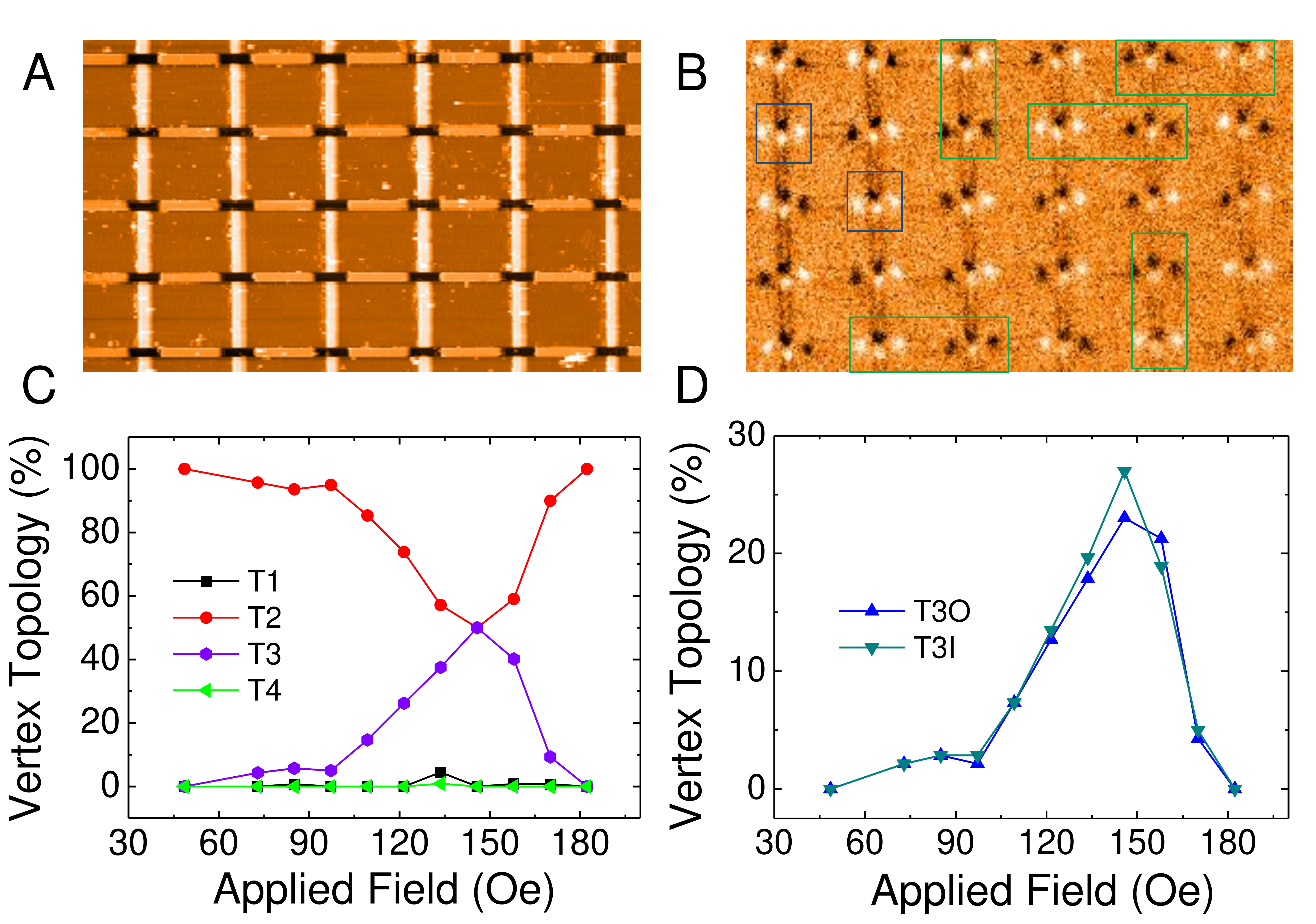}
    \caption{A - Example of topography obtained by AFM in a zoomed region of a sample with $a=\sqrt{2}b$, B - magnetic configuration measured by MFM in the same region, showing that monopoles are in general connected in this low aspect ratio rectangular geometry, with majority of pairs highlighted by green lines while few isolated monopoles are highligheted in blue, C - evolution of all topologies measured in function of external magnetic field applied along the high axis (horizontal) and D - Evolution of topologies carrying magnetic charge split in positive monopoles (T3O) and negative monopoles (T3I). It is possible to notice that both quantities evolve almost together, due to expected pairs formation. Small deviations are related to single monopoles arising in the borders.}
    \label{fig:fig1}
\end{figure}
\begin{figure*}[!hbt]
    \centering
    \includegraphics[scale=0.13]{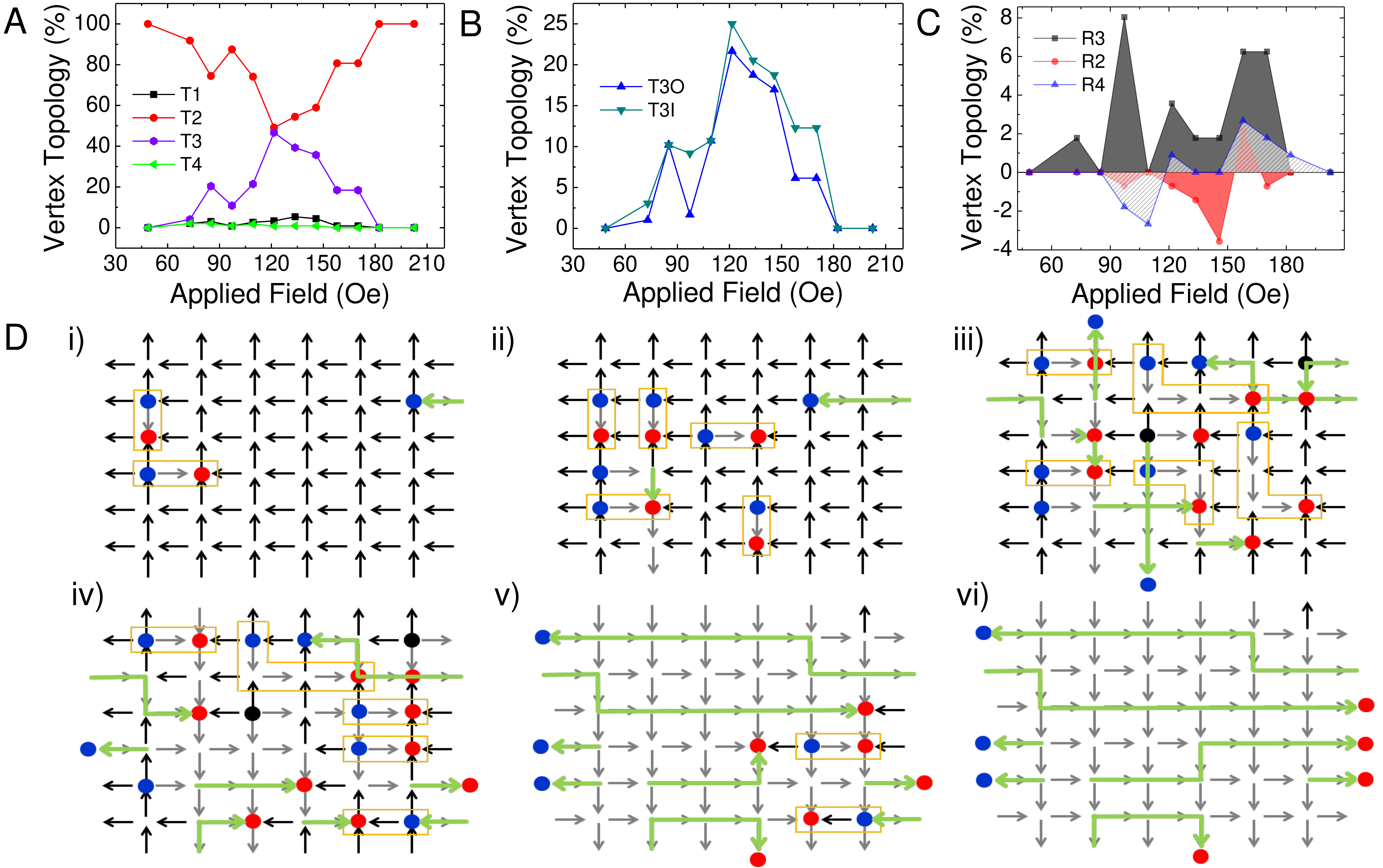}
    \caption{A - Evolution of topologies in function of external field applied in horizontal direction in a rectangular lattice with $a=\sqrt{3}b$. Evolution of different monopoles T3O and T3I in the hard axis, C - Comparison between different quantities of positive and negative monopoles evolution for the three geometries investigated and D - Scheme of monopole creation, transport and annihilation observed in a small part of the sample during measurements.}
    \label{fig:fig1}
\end{figure*}
This is important to elucidate the presence (or absence) of energetic strings and how their tension impacts monopole dynamics. Actually, our main finding relies on the fact that the suitable choice of $a=\sqrt3 b$ provides magnetic monopole to move practically free, since $R3$ array renders a framework with non-energetic strings.

The topography of the $R2$ sample, obtained by atomic force microscopy (AFM), is shown in Figure 2A, while its magnetic signal is depicted in Figure 2B (MFM image has been obtained at remanence after saturation by an applied field along $x$ axis of $180 \rm {Oe}$). Figure 2C shows how vertex population percentage varies with the applied field. Namely, note that, at saturation, essentially $T2$ topology is observed, while at intermediate field $T3$ vertices carrying monopoles also show up, reaching its peak of $\sim 40 \%$, around $150 \rm {Oe}$. In Figures 2D we have focused on the dependence of $T3O$ and $T3I$ vertices evolution in function of the external magnetic field applied along the $x$ direction. Let us recall that the appearance of such vertices is a clear evidence of magnetic monopole pair emerging in the system, while its diminishing indicates pair annihilation. Along $x$-direction, the field is smoothly increased from $40 \rm {Oe}$ to $180 \rm {Oe}$, while $T3$ topology shows up appreciably in the range from $\sim 110 \rm {Oe}$ to $170 \rm {Oe}$, with peak population $\sim 40 \%$ around $150 \rm {Oe}$. The main fact that we would like to point here is the approximate equality in the T3O and T3I quantities during evolution under the external field. Some small deviations could be attributed to monopoles pinned in the borders, as in the example showed in Figure 2B. Such monopole pairing during transport is indeed expected, once in this $a=\sqrt{2}b$ system the string energy is high in both main axis. \cite{nascimento2012confinement}.

In Figure 3A, it is shown how vertices percentage varies with field applied along $x$-axis of $R3$ array. As a whole, we realize a couple of differences with respect to $R2$ array: 1) monopole maximum population is slightly higher, $\sim 50\%$, and it occurs at lower field, $\sim 120 {\rm Oe}$, and; 2) although in less number, $T1$ is now persistent from around $110 \rm {Oe}$ to $160 \rm {Oe}$, which comes to be an experimental evidence of degeneracy between it and $T2$ vertex. 

Actually, a remarkable distinction between $R3$ and any other rectangular arrangement is realized whenever we analyze the relative population of $T3O$ and $T3I$ vertices. Indeed, once a number of magnetic monopole pairs is created in the system, such vertices should show up equally populated. Any mismatch between $T3O$ and $T3I$ number of vertices should then be faced as the appearance of unpaired monopoles. In Figure 3B are depicted $T3O$ and  $T3I$ vertex population as a function of applied field along $x$ direction. A clear mismatch between such populations may be realized, evidencing a considerable number of unpaired monopoles presented in this sample. Indeed, a comparison of the  difference between positive and negative monopoles evolution for the three rectangular geometries investigated is present in Figure 3C. The results show systematic and large differences for R3 geometry, while in the others differences fluctuate around zero and such fluctuation is attributed to monopoles pinning at the borders.

In order to observe the monopole transport in such a very low energetic string framework, we have investigated the evolving of vertex topology in a sample region with $6 \times 6$ vertices, Figure 3D (i)-(vi).\ After a small magnetic field of 75 Oe applied in the horizontal direction, pairs and single monopoles are nucleated, as depicted in Figure 3D(i). By increasing the field to 85 Oe in Figure 3D(ii), more pairs are generated and free monopoles start to move. For little higher fields of 95 Oe and 125 Oe several mechanisms can be observed in Figures 3D(iii) and 3D(iv), as monopole pair extension, annihilation, unpairing for free movement of its components and movement of free monopoles, as well. Finally, in Figures 3D(v) and 3D(vi) acquisted under fields of 170 Oe and 205 Oe, respectively, it is possible to see the last monopole pairs annihilation and the path left by the free monopoles which have crossed the array. With these results we realize that by stretching the square ASI array, in a configuration with higher equidistance possible to a two-dimensional system, one can observe monopole freedom over a background composed by degenerate ground state T1, monopole pairs, dipolar excitations T2 and even high energetic T4 configuration at vertex.

\begin{figure}[!hbt]
    \centering
    \includegraphics[scale=0.13]{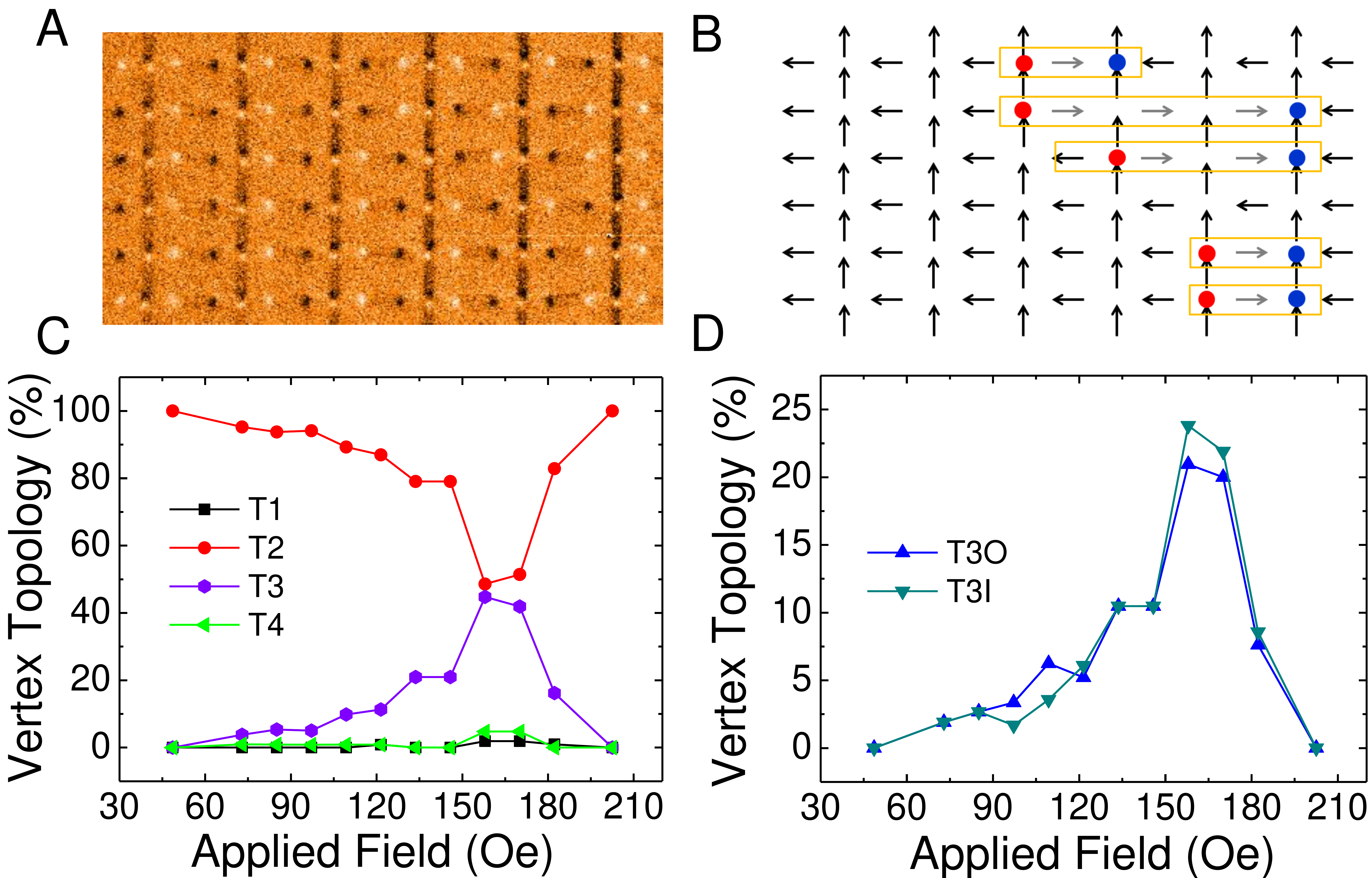}
    \caption{A - Zoom view of MFM measurements performed in a rectangular lattice with $a=\sqrt{4}b$, B - Scheme observed with domain walls formation between monopole pairs, C - Topologies evolution in function of external magnetic field applied in the horizontal direction and D - Evolution of different monopoles T3O and T3I in function of magnetic field in x-axis.}
    \label{fig:fig1}
\end{figure}

Finally, results presented in Figure 4A and 4B confirm the ferromagnetic character expected for the R4 array. Due to the non-interaction among neighbor nanomagnetic stripes, magnetic monopoles evolve forming domain walls between them, in a way similar to that observed for unidirectional ASI \cite{goncalves2019tuning, loreto2015emergence}. Indeed, this may be realized by observing that only $T3$-type vertex takes place in Figure 4C and that monopoles pair move bound by energetic strings, as illustrated in Figures 4D.

In conclusion, we have for the first time presented spin correlation calculations from theoretical investigation of degeneracy in two-dimensional ASI, by dipole Monte Carlo simulation, and experimentally demonstrated such degeneracy in direct measurements of monopole creation, transport and annihilation, during evolution under external magnetic field sweep. Our results showing transport of free monopoles throughout the sample, points the rectangular ASI as the most simple and efficient system to be applied in future magnetricity devices.

\bibliography{reference}

\end{document}